\documentclass[10pt,twocolumn,letterpaper]{article}

\usepackage{wacv}              

\usepackage[accsupp]{axessibility}  
\usepackage{cite}
\usepackage{amsmath,amssymb,amsfonts}
\usepackage{graphicx, overpic} \graphicspath{ {figures_wacv/} }
\usepackage{microtype}
\usepackage{makecell}
\usepackage{multirow}
\usepackage{comment}
\usepackage{hhline}
\usepackage{booktabs}
\usepackage{caption}
\usepackage{arydshln}
\usepackage{tikz}
\usetikzlibrary{positioning,arrows.meta,quotes}
\usetikzlibrary{shapes,decorations}
\usetikzlibrary{bayesnet}
\tikzset{>=latex}
\tikzstyle{plate caption} = [caption, node distance=0, inner sep=0pt,
below left=5pt and 0pt of #1.south]
\usepackage{array}

\usepackage[pagebackref,breaklinks,colorlinks]{hyperref}

\usepackage[capitalize]{cleveref}
\crefname{section}{Sec.}{Secs.}
\Crefname{section}{Section}{Sections}
\Crefname{table}{Table}{Tables}
\crefname{table}{Tab.}{Tabs.}

\def\wacvPaperID{1086} 
\def\confName{WACV}
\def\confYear{2024}

\begin{document}

\title{CSAM: A 2.5D Cross-Slice Attention Module for\\ Anisotropic Volumetric Medical Image Segmentation}

\author{Alex Ling Yu Hung
\quad Haoxin Zheng
\quad Kai Zhao
\quad Xiaoxi Du
\quad Kaifeng Pang
\quad Qi Miao\\
\quad Steven S. Raman
\quad Demetri Terzopoulos
\quad Kyunghyun Sung\\
University of California, Los Angeles\\
}
\maketitle

\begin{abstract}
   A large portion of volumetric medical data, especially magnetic resonance imaging (MRI) data, is anisotropic, as the through-plane resolution is typically much lower than the in-plane resolution. Both 3D and purely 2D deep learning-based segmentation methods are deficient in dealing with such volumetric data since the performance of 3D methods suffers when confronting anisotropic data, and 2D methods disregard crucial volumetric information. Insufficient work has been done on 2.5D methods, in which 2D convolution is mainly used in concert with volumetric information. These models focus on learning the relationship across slices, but typically have many parameters to train. We offer a Cross-Slice Attention Module (CSAM) with minimal trainable parameters, which captures information across all the slices in the volume by applying semantic, positional, and slice attention on deep feature maps at different scales. Our extensive experiments using different network architectures and tasks demonstrate the usefulness and generalizability of CSAM. Associated code is available at \url{https://github.com/aL3x-O-o-Hung/CSAM}.
\end{abstract}

\section{Introduction}

Deep learning (DL) based algorithms have advanced computational medical image analysis by virtue of their superior performance relative to conventional approaches~\cite{litjens2017survey,bakator2018deep}. In particular, convolutional neural networks (CNNs), such as U-Net~\cite{ronneberger2015u}, U-Net++~\cite{zhou2018unet++}, and MSU-Net~\cite{su2021msu}, have achieved state-of-the-art 2D medical image segmentation results. 

Radiologists typically analyze tomographic images as volumes, attending to multiple image slices to make accurate medical decisions. 
2D image segmentation methods, including the aforecited ones, disregard information across slices. Methods using 3D convolutions~\cite{milletari2016v,Chen2018VoxResNet} perform volumetric analysis, but their performance is limited when confronted by anisotropic image data~\cite{jia20193d,isensee2018nnu,isensee2017automatic,hung2022cat}. For example, in T2-weighted MRI the through-plane resolution (3--6\,mm) is typically one-third to one-tenth of the in-plane resolution (0.3--1.0\,mm)~\cite{liu2022evaluation}.  
Furthermore, the gaps between slices and the through-plane resolution can differ for different organs, so one must either upsample the volumes or customize the deep networks to specific datasets and applications. 

Other methods
have been proposed that use only slice-based 2D convolution, but incorporate volumetric information via the 2D feature maps, instead of directly using 3D convolution on the volume. Zhang et al.~\cite{zhang2019multiple} and Han et al.~\cite{han2021liver} stacked nearby slices together as the input to networks, while Yu et al.~\cite{yu2019thickened} and Wang et al.~\cite{wang2019volumetric} combined the input images in the volume in more sophisticated ways. Attention-based methods, such as RsaNet~\cite{zhang2019rsanet}, SAU-Net~\cite{zhang2020sau}, AFTer-U-Net~\cite{yan2022after}, SATr~\cite{li2022satr}, that of Guo and Terzopoulos~\cite{guo2021transformer} (hereafter denoted GT-Net), and our CAT-Net~\cite{hung2022cat}, use cross-slice attention mechanisms to learn the relationship between different slices. Although these methods are promising, it is hard to know the optimal number of slices to use in the input stack or include in the attention, and it is difficult to train a large number of parameters in the Transformer blocks. Despite the challenges, this category of methods, loosely called 2.5D methods, appears to be a better at dealing with anisotropic volumetric medical images. We will provide a formal definition of 2.5D medical image segmentation elsewhere in this paper. 

Instead of using a Transformer network, we propose a Cross-Slice Attention Module (CSAM) for 2.5D medical image segmentation, which significantly reduces the number of trainable parameters relative to other 2.5D methods. Unlike 2.5D segmentation methods that necessitate determining the optimal number of neighboring slices to include, CSAM captures the global cross-slice information across all the slices through its attention mechanism on multi-scale deep feature maps that incorporates the cross-slice semantic and positional attention as well as the importance of the feature maps from different slices. Additionally, we model the uncertainty across slices to better regularize the segmentation. 

The main contributions of this paper are (1) a formal definition of 2.5D medical image segmentation models, (2) a new 2.5D cross-slice attention mechanism that greatly reduces the number of parameters compared to the current state-of-the-art 2.5D models, (3) the CSAM that is conveniently insertable into existing 2D CNN-based networks to enable volumetric image segmentation, (4) incorporation of our CSAMs into different backbone networks, and (5) extensive validation studies on prostate, placenta, and cardiac MRI segmentation, demonstrating improved model performance over corresponding 2D, 3D, and the previous state-of-the-art 2.5D methods.

\section{Related Work}

The U-Net~\cite{ronneberger2015u} encoder-decoder architecture with skip connections, which preserve detailed, high-resolution, and semantic information, has revolutionized medical image segmentation as countless published models are based on or extend this architecture. A fruitful research avenue focuses on combining U-Net with other modules. ResU-Net~\cite{khanna2020deep} introduced the residual connections from ResNet~\cite{he2016deep}, Rundo et al.~\cite{rundo2019use} incorporated the squeeze and excitation (SE) module~\cite{hu2018squeeze}, and Oktay et al.~\cite{Oktay2018AttUNet} injected the widely used attention mechanism~\cite{vaswani2017attention} into U-Net. Other U-Net variants have also emerged. nnU-Net~\cite{isensee2018nnu} replaced the batch normalization~\cite{ioffe2015batch} and rectified linear activation unit (ReLU) with instance normalization~\cite{ulyanov2016instance} and leaky ReLU. To better capture the fine-grained details of the object of interest, U-Net++ uses more nested and dense skip connections between the encoder and decoder. MSU-Net~\cite{su2021msu} added convolutions with different kernel sizes to capture multi-scale information. 

Although 2D methods work reasonably well, they do not consider volume images as a whole since they only analyze the image data slice by slice, neglecting information from other slices. U-Net-inspired 3D networks~\cite{cciccek20163d,milletari2016v,Chen2018VoxResNet} are more effective than 2D U-Net on 3D volumes that have similar cross-voxel distances in all three dimensions and they have been applied to various tasks~\cite{qamar2020variant,xiao2020segmentation,chang2018brain,wang2018two,wang2019deeply}. 

Transformer-based medical image segmentation models are increasing in popularity. They are usually based on CNNs also due to the fact that it is hard to train a pure Transformer model on limited data~\cite{valanarasu2021medical,petit2021u,peiris2022robust,xu2021levit,cao2021swin,hatamizadeh2022unetr}. 

In terms of attention, the squeeze-and-excitation (SE) block~\cite{hu2018squeeze} was proposed to explicitly model the interdependencies between the channels of its convolutional features. Beyond channel-wise attention, the CBAM~\cite{woo2018cbam} models both the channel and spatial attention, which were combined into a 3D attention map in the bottleneck attention module (BAM)~\cite{park2018bam}. Self-attention, cross-attention, and their variants have also been heavily used, especially in vision Transformers~\cite{Oktay2018AttUNet,dosovitskiy2020image,petit2021u,valanarasu2021medical,cao2021swin}. However, limited research has addressed attention between slices, particularly the concept of using attention to mimic clinical decision-making. RsaNet~\cite{zhang2019rsanet} incorporates attention in all three directions in 3D networks instead of 2D networks. SAU-Net~\cite{zhang2020sau} learns attention between different slices of the final feature map but does not learn the whole semantic information. AFTer-U-Net~\cite{yan2022after} applies axial attention to fuse axial information across different slices, but a hyperparameter specifies how many neighboring slices to include in the attention. SATr~\cite{li2022satr} uses a 2D encoder to encode the slice of interest as well as the upper and lower slices and then fuses the encoded feature maps 3D convolution while calculating the attention map with a Transformer block, but a hyperparameter specifies the number of upper and lower slices. GT-Net~\cite{guo2021transformer} applies the attention mechanism in the skip connection at the bottleneck layer of a U-Net. Our CAT-Net~\cite{hung2022cat} uses the cross-slice information at different scales of the deep networks, allowing the model to learn more comprehensive volumetric cross-slice information. Its performance is superior to the other 2.5D methods, but the model requires the training of a large number of parameters.

\section{Defining 2.5D Segmentation}

We denote $x\in\mathbb{R}^{l\times c_0\times h_0\times w_0}$ as the input volumetric image, where $l$ is the total number of slices in the volume, $c_0$ is the number of channels per slice, and $h_0$ and $w_0$ are the height and width of the inputs. In most cases, there is only one channel per slice; i.e., $c_0=1$, for a single modality. However, there may be multiple channels with multi-parametric MRI, such as T1 weighted imaging, T2 weighted imaging, apparent diffusion coefficient (ADC) map, etc., or for multi-modality imaging, such as combined MRI and Computed Tomography (CT). We also denote a pure 2D encoder as $E$ and decoder as $D$. 

Conventionally, researchers stack nearby slices as additional channels in the input to 2D segmentation networks and segment the middle slice, referring to this as 2.5D segmentation~\cite{zhang2019multiple,han2021liver}; i.e.,
\begin{equation}
    y=D(E(F_{\text{cat}}(x))),
    \label{equation:stack}
\end{equation}
where $y$ is the segmentation mask and $F_{\text{cat}}$ denotes the concatenation of nearby slices. Methods that use other ways to combine the input images in the volume, such as those of Yu et al.~\cite{yu2019thickened} and Wang et al.~\cite{wang2019volumetric}, can also be expressed in the form of (\ref{equation:stack}).

\begin{figure*}
    \centering
    \includegraphics[width=1\textwidth]{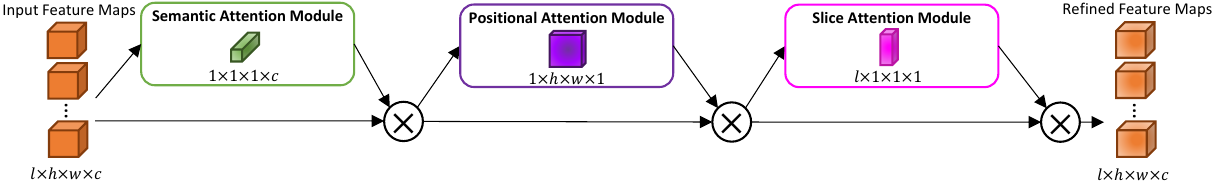}
     \caption{The CSAM architecture. The input feature maps from different slices are fed into the module concurrently, where they go through three sequential sub-modules: the semantic, positional, and slice attention modules.}
     \label{fig:c2bam}
 \end{figure*}

By contrast, recent models~\cite{zhang2019rsanet,yan2022after,li2022satr,guo2021transformer,hung2022cat} incorporate different cross-slice attention mechanisms on deep feature maps between the encoder and decoder, expressible as 
\begin{equation}
    y=D(F_{\text{attention}}(E(x))),
    \label{equation:cat}
\end{equation}
where $F_{\text{attention}}$ is the cross-slice attention operation between the encoder and decoder. Most of them do not explicitly claim to be 2.5D methods, but we regard them as such since they consider the relationship between slices. 

Additionally, SAU-Net~\cite{zhang2020sau} uses a cross-slice attention mechanism after the decoder to learn the cross-slice relationship, which can be expressed as 
\begin{equation}
    y=F_{\text{attention}}(D(E(x))),
    \label{equation:sau}
\end{equation}
where $F_{\text{attention}}$ is the cross-slice attention operation following the decoder. 

Since there has been no formal consensus on the 2.5D segmentation approach, we define the general form of 2.5D segmentation as 
\begin{equation}
    y=F_{\text{post}}(D(F_{\text{mid}}(E(F_{\text{pre}}(x))))),
    \label{equation:general}
\end{equation}
where $F_{\text{pre}}$ is a function applied to the input volume, $F_{\text{mid}}$ is the operation between the 2D encoder and decoder, and $F_{\text{post}}$ is the operation after the decoder. Thus,  \textit{for any 2D encoder $E$ and decoder $D$, when at least one of $F_{\text{pre}}$, $F_{\text{mid}}$, or $F_{\text{post}}$ in (\ref{equation:general}) involves operations between different slices, we categorize the segmentation model as being 2.5D.} Note that, (\ref{equation:general}) reduces to (\ref{equation:stack}) when $F_{\text{pre}}$ represents a concatenation of nearby slices while $F_{\text{mid}}= F_{\text{post}}=I$ (i.e., identity functions), it reduces to (\ref{equation:cat}) when $F_{\text{mid}}$ represents cross-slice attention between the encoder and decoder while $F_{\text{pre}}= F_{\text{post}}=I$, and it reduces to (\ref{equation:sau}) when $F_{\text{post}}$ becomes the cross-slice attention mechanism while $F_{\text{pre}}=F_{\text{mid}}=I$. 

\section{The Cross-Slice Attention Module}
\subsection{Overview}

The CSAM incorporates information from other slices within the image volume to perform semantic, positional, and slice attention in 2.5D image segmentation. The CSAM inputs the feature maps from all slices at different scales and outputs the refined feature maps. As shown in Figure~\ref{fig:c2bam}, given feature maps $F\in\mathbb{R}^{l\times h\times w\times c}$ from all the slices in the volume, where $l$ is the total number of slices in the volume, $h$ and $w$ determine the size of the feature maps, and $c$ is the number of channels, the CSAM sequentially calculates the semantic $M_{\text{semantic}}\in\mathbb{R}^{1\times1\times1\times c}$, positional $M_{\text{positional}}\in\mathbb{R}^{1\times h\times w\times1}$, and slice $M_{\text{slice}}\in\mathbb{R}^{l\times1\times1\times1}$ attention maps, which are used to weigh the input feature maps $F$, thus obtaining the refined feature maps $F'$ as follows:
\begin{align}
    F_1&=M_{\text{semantic}}(F)\otimes F,\quad \\
    F_2&=M_{\text{positional}}(F_1)\otimes F_1,\quad \\
    F'&=M_{\text{slice}}(F_2)\otimes F_2,
\end{align}
where $F_1,F_2\in\mathbb{R}^{l\times h\times w\times c}$ are intermediate results and $\otimes$ denotes element-wise multiplication. We perform broadcasts accordingly during the multiplication. 

\subsection{Semantic Attention}

\begin{figure}
    \centering
    \includegraphics[width=1\linewidth]{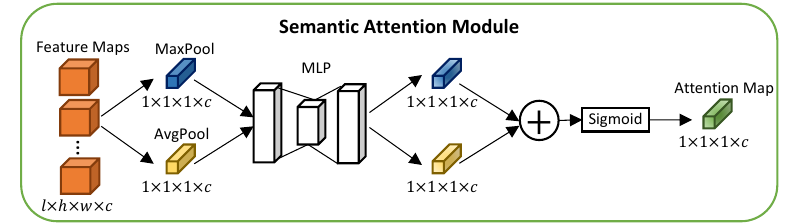}
     \caption{The semantic attention.}
     \label{fig:submodules-semantic}
\end{figure}

The semantic attention module (Figure~\ref{fig:submodules-semantic}) takes feature maps $F$ as input and produces a semantic attention map $M_{\text{semantic}}$, which focuses on what semantic information is important among all the 2D feature maps of the volume. Unlike conventional 2D approaches, our semantic attention map is the same for all 2D feature maps in the volume. 
To compute the semantic feature map, the $l$, $h$, and $w$ dimensions are squeezed by max and average pooling as these pooling methods gather different information about the features for better semantic attention~\cite{woo2018cbam}. 
First, global max and average pooling are performed on the $l$, $h$ and $w$ dimensions. Second, the max and averagepooled features are separately fed into the same multi-layer perceptron (MLP). Finally, the sum of the two outputs is fed into a sigmoid function to generate the semantic attention map. The operations can be written as
\begin{equation}
    M_{\text{semantic}}(F)=\sigma(\text{MLP}(\text{MP}(F))+\text{MLP}(\text{AP}(F))),
\end{equation}
where $\sigma(\cdot)$ is the sigmoid function and $\text{MP}(\cdot)$ and $\text{AP}(\cdot)$ denote max and average pooling, respectively. Note that the $l\times h\times w\times c$ input feature maps are pooled to $1\times1\times1\times c$; i.e., the semantic attention weights are the same on a single channel at different spatial locations within the slice and between slices.

\subsection{Positional Attention}

\begin{figure}
    \centering
    \includegraphics[width=1\linewidth]{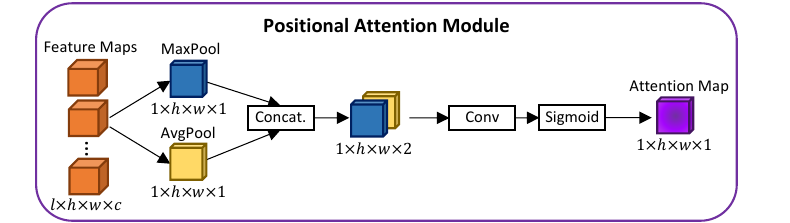}
     \caption{The positional attention.}
     \label{fig:submodules-positional}
\end{figure}

The positional attention module (Figure~\ref{fig:submodules-positional}) focuses on where the important information is on each slice of the input volume. It takes feature maps $F_1$ as inputs and outputs a positional attention map $M_{\text{positional}}$. The features maps are max and averagepooled across $l$ and $c$ dimensions to two $1\times h\times w\times1$ feature maps that are subsequently concatenated and then subjected to a convolution operation and a sigmoid function, as follows:
\begin{equation}
    M_{\text{positional}}(F_1)=\sigma(\text{conv}[\text{MP}(F_1);\text{AP}(F_1)]),
\end{equation}
where $\text{conv}[.;.]$ denotes a convolution operation with the semicolon denoting concatenation.

\subsection{Slice Attention}

\begin{figure}
    \centering
    \includegraphics[width=1\linewidth]{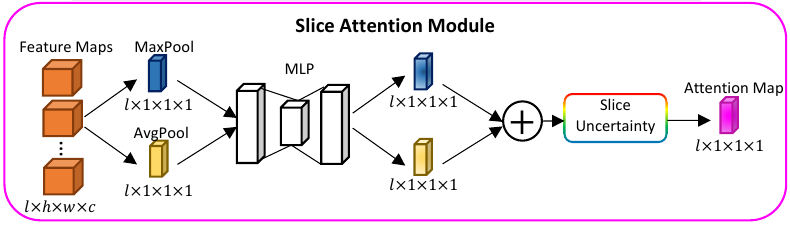}
     \caption{The slice attention.}
     \label{fig:submodules-slice}
\end{figure}

The slice attention module (Figure~\ref{fig:submodules-slice}) is designed to compute which slices are more likely to contain important information relevant to the intended task and suppress features that might lead to false positive predictions on particular slices; e.g., the module should infer that it is less likely to have positive prostate segmentation in the first and last slice of the volume than in the middle of the volume. Furthermore, the slice uncertainty block within the slice attention module can capture the aleatoric uncertainty inherently stemming from the input data while also acting as a form of regularization to improve performance.

The slice attention module first max and average pools the input feature maps and feeds them separately into the same MLP. However, here the input feature maps are pooled in $h$, $w$, and $c$ dimensions, making the pooled features $l\times1\times1\times1$. Before generating the slice attention map $M_\text{slice}$, the outputs of the MLP are added together and fed into a slice uncertainty block, which captures the uncertainty caused by the partial volume effects~\cite{ballester2002estimation} or the difference in length between the slice thickness and the spatial resolution. The slice attention on different slices should be correlated, and the correlation should be stronger on nearby slices. Therefore, 
we use a low-rank Gaussian distribution $\mathcal{N}(\mu,\Sigma)$ to model the slice uncertainty so that the three components describing the distribution, namely the mean $\mu$, covariance factor $P$, and the covariance diagonal $D$ can all be computed efficiently while approximating the underlying Gaussian distribution. The input to the slice uncertainty block is $V\in\mathbb{R}^{l\times1}$, which is the reshaped tensor of the summation of two $l\times1\times1\times1$ tensors. Three linear operations are performed on this tensor: 
\begin{align}
    \mu' = W_\mu V,\qquad 
    P' = W_P V,\qquad 
    D' = W_D V,
\end{align}
where $W_\mu\in\mathbb{R}^{l\times l}$, $W_D\in\mathbb{R}^{l\times l}$, $W_P\in\mathbb{R}^{(lr)\times l}$, and $r$ is the rank of the parameterization. The covariance matrix $\Sigma$ of the distribution is calculated as
\begin{equation}
    \Sigma=PP^T+D,
\end{equation}
where $P\in\mathbb{R}^{l\times r}$ is the reshaped tensor of $P'$ and $D\in\mathbb{R}^{l\times l}$ is a diagonal matrix with elements of $D'$ on the diagonal. The mean of the low-rank Gaussian distribution is $\mu\in\mathbb{R}^l$, which is the reshaped tensor of $\mu'$. Next, a vector $z$ is sampled from the distribution and passed through a sigmoid function to obtain the un-reshaped slice attention map
\begin{equation}
    M_{\text{slice}}'=\sigma(z); \qquad z\sim\mathcal{N}(\mu,\Sigma).
\end{equation}
The final output $ M_{\text{slice}}\in\mathbb{R}^{l\times1\times1\times1}$ is the reshaped $M_{\text{slice}}'$. 

\subsection{Implementation on Different Backbones}

CSAMs are incorporated into U-Net-like backbone networks with skip connections between the encoder $E$ and decoder $D$. We define $X_{i,j}$ as a basic convolution block of the backbone network and $x_{i,j}$ as the output of convolution block $X_{i,j}$. $E$ is made up of the convolution blocks $X_{i,0}$, and all the other convolution blocks make up $D$. 

Taking in a tensor $x\in\mathbb{R}^{l\times c_0\times h_0\times w_0}$ with $l$, $c_0$, $h_0$, and $w_0$ being the total number of slices in the volume, the number of channels, and the height and width of the inputs, respectively, $E$ treats $l$ as the batch dimension in conventional 2D approaches and outputs a set of encoded feature maps
$\{x_{i,0}\}_{0\leq i<L}=E(x)$,
where $x_{i,0}\in\mathbb{R}^{l\times c_i\times h_i\times w_i}$ is the output of each convolution block and $L$ is the number of layers in the network.

\begin{figure}
    \centering
    \scalebox{0.5}{
    \begin{tikzpicture}
\node [circle,draw=black,fill=black!30,inner sep=0pt,minimum size=1cm] (x00) at (-8,-2) {$X_{0,0}$};
\node [circle,draw=black,fill=black!30,inner sep=0pt,minimum size=1cm] (x10) at (-7,-3.5) {$X_{1,0}$};
\node [circle,draw=black,fill=black!30,inner sep=0pt,minimum size=1cm] (x20) at (-6,-5) {$X_{2,0}$};
\node [circle,draw=black,fill=black!30,inner sep=0pt,minimum size=1cm] (x30) at (-5,-6.5) {$X_{3,0}$};
\node [circle,draw=black,fill=black!30,inner sep=0pt,minimum size=1cm] (x40) at (-4,-8) {$X_{4,0}$};

\node [rectangle,draw=black,fill=white,inner sep=0pt,minimum size=1cm] (isim0) at (-3,-2) {$CSAM_0$};
\node [rectangle,draw=black,fill=white,inner sep=0pt,minimum size=1cm] (isim1) at (-3,-3.5) {$CSAM_1$};
\node [rectangle,draw=black,fill=white,inner sep=0pt,minimum size=1cm] (isim2) at (-3,-5) {$CSAM_2$};
\node [rectangle,draw=black,fill=white,inner sep=0pt,minimum size=1cm] (isim3) at (-3,-6.5) {$CSAM_3$};
\node [rectangle,draw=black,fill=white,inner sep=0pt,minimum size=1cm] (isim4) at (-2,-8) {$CSAM_4$};

\node [circle,draw=black,fill=white,inner sep=0pt,minimum size=1cm] (x31) at (-1,-6.5) {$X_{3,1}$};

\node [circle,draw=black,fill=white,inner sep=0pt,minimum size=1cm] (x21) at (0,-5) {$X_{2,1}$};

\node [circle,draw=black,fill=white,inner sep=0pt,minimum size=1cm] (x11) at (1,-3.5) {$X_{1,1}$};

\node [circle,draw=black,fill=white,inner sep=0pt,minimum size=1cm] (x01) at (2,-2) {$X_{0,1}$};

\path [draw,->] (x00)edge[] (isim0);
\path [draw,->] (isim0)edge[densely dotted] (x01);

\path [draw,->] (x10)edge[] (isim1);
\path [draw,->] (isim1)edge[densely dotted] (x11);

\path [draw,->] (x20)edge[] (isim2);
\path [draw,->] (isim2)edge[densely dotted] (x21);

\path [draw,->] (x30)edge[] (isim3);
\path [draw,->] (isim3)edge[densely dotted] (x31);

\path [draw,->] (x40)edge[] (isim4);

\path [draw,->] (isim4)edge[] (x31);

\path [draw,->] (x00)edge[] (x10);
\path [draw,->] (x10)edge[] (x20);
\path [draw,->] (x20)edge[] (x30);
\path [draw,->] (x30)edge[] (x40);

\path [draw,->] (x21)edge[] (x11);
\path [draw,->] (x31)edge[] (x21);
\path [draw,->] (x11)edge[] (x01);
\path [draw,->] (x01)edge[] (3.5,-2);
\coordinate (temp) at (-9.5,-2);
\draw[->] (temp) -- (x00);
\end{tikzpicture}}

~~~~~~~~~~~~~~~~~~~~~~~~~~

\scalebox{0.5}{
\begin{tikzpicture}
\node [circle,draw=black,fill=black!30,inner sep=0pt,minimum size=1cm] (x00) at (-8,-2) {$X_{0,0}$};
\node [circle,draw=black,fill=black!30,inner sep=0pt,minimum size=1cm] (x10) at (-7,-3.5) {$X_{1,0}$};
\node [circle,draw=black,fill=black!30,inner sep=0pt,minimum size=1cm] (x20) at (-6,-5) {$X_{2,0}$};
\node [circle,draw=black,fill=black!30,inner sep=0pt,minimum size=1cm] (x30) at (-5,-6.5) {$X_{3,0}$};
\node [circle,draw=black,fill=black!30,inner sep=0pt,minimum size=1cm] (x40) at (-4,-8) {$X_{4,0}$};

\node [rectangle,draw=black,fill=white,inner sep=0pt,minimum size=1cm] (isim0) at (-6,-2) {$CSAM_0$};
\node [rectangle,draw=black,fill=white,inner sep=0pt,minimum size=1cm] (isim1) at (-5,-3.5) {$CSAM_1$};
\node [rectangle,draw=black,fill=white,inner sep=0pt,minimum size=1cm] (isim2) at (-4,-5) {$CSAM_2$};
\node [rectangle,draw=black,fill=white,inner sep=0pt,minimum size=1cm] (isim3) at (-3,-6.5) {$CSAM_3$};
\node [rectangle,draw=black,fill=white,inner sep=0pt,minimum size=1cm] (isim4) at (-2,-8) {$CSAM_4$};

\node [circle,draw=black,fill=white,inner sep=0pt,minimum size=1cm] (x01) at (-4,-2) {$X_{0,1}$};
\node [circle,draw=black,fill=white,inner sep=0pt,minimum size=1cm] (x11) at (-3,-3.5) {$X_{1,1}$};
\node [circle,draw=black,fill=white,inner sep=0pt,minimum size=1cm] (x21) at (-2,-5) {$X_{2,1}$};
\node [circle,draw=black,fill=white,inner sep=0pt,minimum size=1cm] (x31) at (-1,-6.5) {$X_{3,1}$};

\node [circle,draw=black,fill=white,inner sep=0pt,minimum size=1cm] (x02) at (-2,-2) {$X_{0,2}$};
\node [circle,draw=black,fill=white,inner sep=0pt,minimum size=1cm] (x12) at (-1,-3.5) {$X_{1,2}$};
\node [circle,draw=black,fill=white,inner sep=0pt,minimum size=1cm] (x22) at (0,-5) {$X_{2,2}$};

\node [circle,draw=black,fill=white,inner sep=0pt,minimum size=1cm] (x03) at (0,-2) {$X_{0,3}$};
\node [circle,draw=black,fill=white,inner sep=0pt,minimum size=1cm] (x13) at (1,-3.5) {$X_{1,3}$};

\node [circle,draw=black,fill=white,inner sep=0pt,minimum size=1cm] (x04) at (2,-2) {$X_{0,4}$};

\path [draw,->] (x00)edge[] (isim0);
\path [draw,->] (isim0)edge[densely dotted] (x01);
\path [draw,->] (x01)edge[densely dotted] (x02);
\path [draw,->] (x02)edge[densely dotted] (x03);
\path [draw,->] (x03)edge[densely dotted] (x04);

\path [draw,->] (x10)edge[] (isim1);
\path [draw,->] (isim1)edge[densely dotted] (x11);
\path [draw,->] (x11)edge[densely dotted] (x12);
\path [draw,->] (x12)edge[densely dotted] (x13);

\path [draw,->] (x20)edge[] (isim2);
\path [draw,->] (isim2)edge[densely dotted] (x21);
\path [draw,->] (x21)edge[densely dotted] (x22);

\path [draw,->] (x30)edge[] (isim3);
\path [draw,->] (isim3)edge[densely dotted] (x31);

\path [draw,->] (x40)edge[] (isim4);

\path [draw,->] (isim1)edge[] (x01);
\path [draw,->] (isim2)edge[] (x11);
\path [draw,->] (isim3)edge[] (x21);
\path [draw,->] (isim4)edge[] (x31);

\path [draw,->] (x00)edge[] (x10);
\path [draw,->] (x10)edge[] (x20);
\path [draw,->] (x20)edge[] (x30);
\path [draw,->] (x30)edge[] (x40);

\path [draw,->] (x31)edge[] (x22);
\path [draw,->] (x22)edge[] (x13);
\path [draw,->] (x13)edge[] (x04);

\path [draw,->] (x21)edge[] (x12);
\path [draw,->] (x12)edge[] (x03);

\path [draw,->] (x11)edge[] (x02);

\path [draw,->] (isim0) edge[,densely dotted, out=35,in=145,looseness=0.5] (x02);
\path [draw,->] (isim1) edge[,densely dotted, out=35,in=145,looseness=0.5] (x12);
\path [draw,->] (isim2) edge[,densely dotted, out=35,in=145,looseness=0.5] (x22);

\path [draw,->] (isim0) edge[,densely dotted, out=35,in=145,looseness=0.5] (x03);
\path [draw,->] (isim1) edge[,densely dotted, out=35,in=145,looseness=0.5] (x13);

\path [draw,->] (isim0) edge[,densely dotted, out=35,in=145,looseness=0.5] (x04);

\path [draw,->] (x01) edge[,densely dotted, out=35,in=145,looseness=0.5] (x03);
\path [draw,->] (x11) edge[,densely dotted, out=35,in=145,looseness=0.5] (x13);

\path [draw,->] (x01) edge[,densely dotted, out=35,in=145,looseness=0.5] (x04);
\path [draw,->] (x02) edge[,densely dotted, out=35,in=145,looseness=0.5] (x04);

\path [draw,->] (x04)edge[] (3.5,-2);
\coordinate (temp) at (-9.5,-2);
\draw[->] (temp) -- (x00);
\end{tikzpicture}
}
    \caption{Implementation of the CSAM (top) in the nnU-Net~\cite{isensee2018nnu} and MSU-Net~\cite{su2021msu}, and (bottom) in the nnU-Net++~\cite{zhou2018unet++}.  
 The gray circles represent the 2D encoder. The white circles represent the 2D decoder.}
\label{fig:segmentation_implementation}
\end{figure}
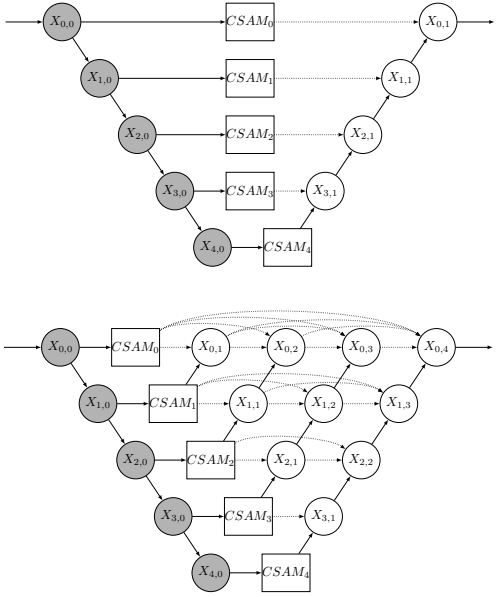

We use $D_i$ to represent layer $i$ of the decoder and $d_i$ as the corresponding output, where $0\leq i<L-1$. We manually define the output of $\text{CSAM}_{L-1}$, which is the deepest CSAM, as $d_{L-1}$; i.e. $d_{L-1}=\text{CSAM}_{L-1}(x_{L-1,0})$. When $0\leq i<L-1$, in a U-Net-based framework, $D_i$ is equivalent to $X_{i,1}$, $d_i$ is the same as $x_{i,1}$, and the final output is $x_{0,1}$, whereas in a U-Net++-based framework, $D_i$ consists of a set of convolution blocks, namely $D_i=\{X_{i,j}\}_{1\leq j<L-i}$ and $d_i=\{x_{i,j}\}_{1\leq j<L-i}$, and the final output is $x_{0,L-1}$. As shown in Figure~\ref{fig:segmentation_implementation}, the deep feature maps $x_{i,0}$ at each encoder layer $i$ are injected into the CSAMs before getting to the decoder:
\begin{equation}
    d_i=D_i(\text{CSAM}_i(x_{i,0}),\mathcal{U}(d_{i+1})),
\end{equation}
where $\text{CSAM}_i$ is associated with layer $i$, and $\mathcal{U}(.)$ denotes the upsampling operation.

\section{Experiments}



\subsection{Implementation and Evaluation Details}

The CSAM was tested on segmentation of prostate, placenta, and cardiac MRI, with nnU-Net~\cite{isensee2018nnu}, MSU-Net~\cite{su2021msu}, and nnU-Net++~\cite{zhou2018unet++} as backbones. It was also evaluated on prostate MRI cancer segmentation. Cross entropy loss was used on prostate zonal and cardiac segmentation, dice loss for placenta segmentation, and focal loss\cite{Kaiming2020FocalLoss} for prostate cancer (PCa) segmentation. 
We used the Adam~\cite{kingma2014adam} optimizer with a learning rate of 0.0001 and weight decay regularization
with the parameter set to $1\times10^{-5}$. Only center crop, horizontal flip, and Gamma transform were performed as augmentation. 

The Dice similarity coefficient (DSC) and relative absolute volume difference (RAVD) were calculated in 3D. Patient-wise classification AUC was calculated for prostate MRI cancer segmentation. For experiments involving cross-validation, we performed statistical analysis with the Mann-Whitney U Test~\cite{nachar2008mann} to compare the distribution of results from our CSAM-based models against those of the competing models. 

\subsection{Prostate MRI Zonal Segmentation}

This dataset comprises 296 patients, including only T2-weighted MRI scans with an in-plane resolution of 0.625\,mm$^2$ and a through-plane resolution of 3\,mm. The prostate transition zone (TZ) and peripheral zone (PZ) were manually annotated by clinical experts as ground truth. We randomized the data and grouped them into five folds for cross-validation. For the training of the 2D and 2.5D models, the input volume size is $128\times128\times20$. To perform multiple downsampling operations in the 3D models, we upsampled the volumes along the z-axis to $128\times128\times80$. We carried out 5-fold cross-validation in this experiment.

Our ablation study was performed on the prostate zonal segmentation using a nnU-Net++-based architecture. As reported in Table~\ref{tab:ablation}, all three sub-modules can improve the segmentation performance, while using attention in all three dimensions achieves the best results with statistical significance in terms of DSC in both the TZ and PZ. Furthermore, we compared model performance including and excluding the slice uncertainty on the same task. From Table~\ref{tab:uncertainty}, we see that compared to the absence of slice uncertainty modeling, both using just the slice attention module and using all three attention modules along with the slice uncertainty can improve performance by a good amount.

\begin{table} \setlength{\tabcolsep}{4pt}
\centering
\begin{tabular}{ ccc lll lll }
\toprule
Semantic&Positional&Slice&TZ&PZ\\
\midrule
&&&87.9***&82.4***\\
\checkmark&&&88.0***&83.2***\\
&\checkmark&&88.0***&83.0***\\
&&\checkmark&88.3***&83.7***\\
\checkmark&\checkmark&&88.7***&84.2***\\
\checkmark&&\checkmark&88.5***&83.9***\\
&\checkmark&\checkmark&88.4***&84.0***\\
\checkmark&\checkmark&\checkmark&\bfseries89.9&\bfseries85.2\\
\bottomrule
\end{tabular}
\caption{Ablation study of zonal segmentation (DSC in \%).  Note: in this and subsequent tables, the labels *, **, *** indicate p-values of $\leq0.1$, $\leq0.05$, and $\leq0.01$, respectively. All the experiments involving cross-validation adhere to the same notation.}
\label{tab:ablation}
\end{table}

\begin{table} \setlength{\tabcolsep}{4pt}
\centering
\begin{tabular}{ l lll lll }
\toprule
&TZ &PZ \\
\midrule
Slice without uncertainty&87.6***&82.8***\\
Slice with uncertainty&\bfseries88.3&\bfseries83.7\\
\midrule
All without uncertainty&88.3***&84.7***\\
All with uncertainty&\bfseries89.9&\bfseries85.2\\
\bottomrule
\end{tabular}
\caption{Ablation study of the effect of using slice uncertainty (DSC in \%).}
\label{tab:uncertainty}
\end{table}

\begin{table*} \setlength{\tabcolsep}{4pt}
\centering
\begin{tabular}{ lll lll}
\toprule
& \multicolumn{2}{c}{TZ} & \multicolumn{2}{c}{PZ}& \\
\cmidrule(lr){2-3} \cmidrule(lr){4-5}
 & DSC (\%)$\uparrow$ & RAVD ($\%$)$\downarrow$ & DSC (\%)$\uparrow$ &
RAVD ($\%$)$\downarrow$ &\# Parameters\\
\midrule
nnU-Net~\cite{isensee2018nnu}& 88.5***&23.6***&83.6***&32.8***&138M\\
3D nnU-Net~\cite{isensee2018nnu}&89.0***&22.8***&83.6***&32.1***&264M\\
GT-nnU-Net~\cite{guo2021transformer}&87.7***&25.6***&82.7***&34.6***&186M\\
SAnnU-Net~\cite{zhang2020sau}&87.6***&25.3***&82.3***&35.4***&138M\\
Satr-nnU-Net~\cite{li2022satr}&85.8***&29.8***&79.0***&43.2***&269M\\
CAT-nnU-Net~\cite{hung2022cat} &89.5&\bfseries20.8&85.1**&29.8**&614M\\
CSAM-nnU-Net & \bfseries89.7&21.0&\bfseries85.7&\bfseries29.1&139M\\
\midrule
MSU-Net~\cite{su2021msu}&88.1&24.8**&82.3&33.5*&47.1M\\
3D MSU-Net~\cite{su2021msu}&\bfseries89.3&\bfseries21.8&\bfseries83.6&32.6&293M\\
GT-MSU-Net~\cite{guo2021transformer}&86.1***&28.5***&79.2***&42.0***&55.0M\\
SAMSU-Net~\cite{zhang2020sau}&87.1**&26.5**&81.1**&37.6**&47.1M\\
Satr-MSU-Net~\cite{li2022satr}&84.4***&31.8***&78.8***&41.2***&87.1M\\
CAT-MSU-Net~\cite{hung2022cat}&87.0***&25.3***&81.7*&34.0**&264M\\
CSAM-MSU-Net&89.0&\bfseries21.8&83.1&\bfseries31.8&47.2M\\
\midrule
nnU-Net++~\cite{zhou2018unet++}&87.9***&23.9***&82.4***&34.0**&36.6M\\
3D nnU-Net++~\cite{zhou2018unet++}&88.1***&23.8***&82.2***&34.2***&110M\\
GT-nnU-Net++~\cite{guo2021transformer}&88.6***&23.3***&83.8**&33.0**&57.6M\\
SAnnU-Net++~\cite{zhang2020sau}&85.1***&30.6***&77.7***&44.5***&36.6M\\
Satr-nnU-Net++~\cite{li2022satr}&88.4***&24.1***&83.3***&33.9***&101M\\
CAT-nnU-Net++~\cite{hung2022cat}&\bfseries89.9&\bfseries20.6&\bfseries85.6&\bfseries29.1&398M\\
CSAM-nnU-Net++&\bfseries89.9&\bfseries20.6&85.2&29.5&36.8M\\
\bottomrule
\end{tabular}
\caption{Comparison of CSAM-Net against the corresponding 2D, 3D, and 2.5D models for prostate zonal segmentation.}
\label{tab:prostate_zonal_segmentation}
\end{table*}

To test the generalizability of our CSAM across different backbones, we incorporated it into nnU-Net, MSU-Net, and nnU-Net++, and compared them with the corresponding 2D model, 3D model, and 2.5D GT-Net~\cite{guo2021transformer}, SA-Net~\cite{zhang2020sau}, Satr-Net~\cite{li2022satr}, and CAT-Net~\cite{hung2022cat}.
Table~\ref{tab:prostate_zonal_segmentation} shows that on nnU-Net, CSAMs significantly increased segmentation performance relative to the 2D and 3D nnU-Net for both the TZ and PZ, while the performance is similar for the TZ compared with CAT-nnU-Net. However, CSAM-nnU-Net significantly outperforms CAT-nnU-Net for the segmentation of the PZ, while employing considerably fewer parameters. The other 2.5D models underperformed the 2D nnU-Net model due to the fact that prostate zonal segmentation requires relatively little volumetric information and less carefully designed attention mechanisms can confuse the model. In terms of MSU-Net-based network structures, CSAM-MSU-Net significantly outperforms 2D MSU-Net, GT-MSU-Net, SAMSU-NET, Satr-MSU-Net, and CAT-MSU-Net, while exhibiting similar performance to 3D MSU-Net. In nnU-Net++, CSAMs have similar performance to CAT modules, while outperforming the corresponding 2D and 3D methods. In short, CSAM-Net showed superior performance when using any of nnU-Net, MSU-Net, and nnU-Net++ architectures in prostate zonal segmentation. Furthermore, the performance of CSAM-Net is similar to that of CAT-Net or the corresponding 3D method, but it remains a favorable option due to its much smaller number of parameters. 

We evaluated the slice uncertainty with CSAM-nnU-Net. We investigated the lower quartile, median, and upper quartile of the uncertainty values of the correctly and incorrectly segmented pixels. The upper quartile uncertainty value of correctly segmented pixels was 0, while the incorrectly segmented pixels had a lower quartile of 0.04, a median of 0.09, and an upper quartile of 0.15. This indicates that uncertainty estimation can be used to discriminate between correctly and incorrectly segmented pixels. Furthermore, by examining the pixel-wise segmentation error rates corresponding to each uncertainty value, we found that the correlation between the error rates and the uncertainty values is $r=0.791$.  

\subsection{Prostate MRI Cancer Segmentation}

The dataset consists of 652 patients with T2-weighted imaging, ADC, and high b-value (b=1400) diffusion-weighted imaging (DWI) MRI, with volume sizes of $128\times128\times20$. The in-plane resolution is 0.625\,mm$^2$, and the through-plane resolution is 3\,mm. They were stacked together with the manual zonal segmentation as inputs to the models. 220 patients were confirmed with PCa, while 432 patients had no indication of PCa with biopsy. 5-fold cross-validation was applied in the experiment.

\begin{table}
    \centering
    \begin{tabular}{lll}
    \toprule
    &DSC (\%)$\uparrow$& cls. AUC $\uparrow$\\
    \midrule
         3D nnU-Net~\cite{isensee2018nnu}&37.0**&0.854**\\
         3D Residual U-Net~\cite{Bhalerao2020ResUNet}&39.3**&0.868**\\
         3D SEResU-Net~\cite{hu2018squeeze}&45.1&0.850***\\
         3D Attention U-Net~\cite{Oktay2018AttUNet}&45.5&0.841***\\
         VNet~\cite{milletari2016v}&44.1*&0.846**\\
         VoxResNet~\cite{Chen2018VoxResNet}&47.2&0.868**\\
         UNETR~\cite{hatamizadeh2022unetr}&43.6&0.852**\\
         VTU-Net~\cite{peiris2022robust}&44.3*&0.846***\\
         CSAM-nnU-Net&\bfseries51.0&\bfseries0.942\\
    \bottomrule
    \end{tabular}
    \caption{Results of prostate cancer segmentation.}
    \label{tab:cancer_segmentation}
\end{table}

 \begin{figure}
    \centering
    \includegraphics[width=1\linewidth]{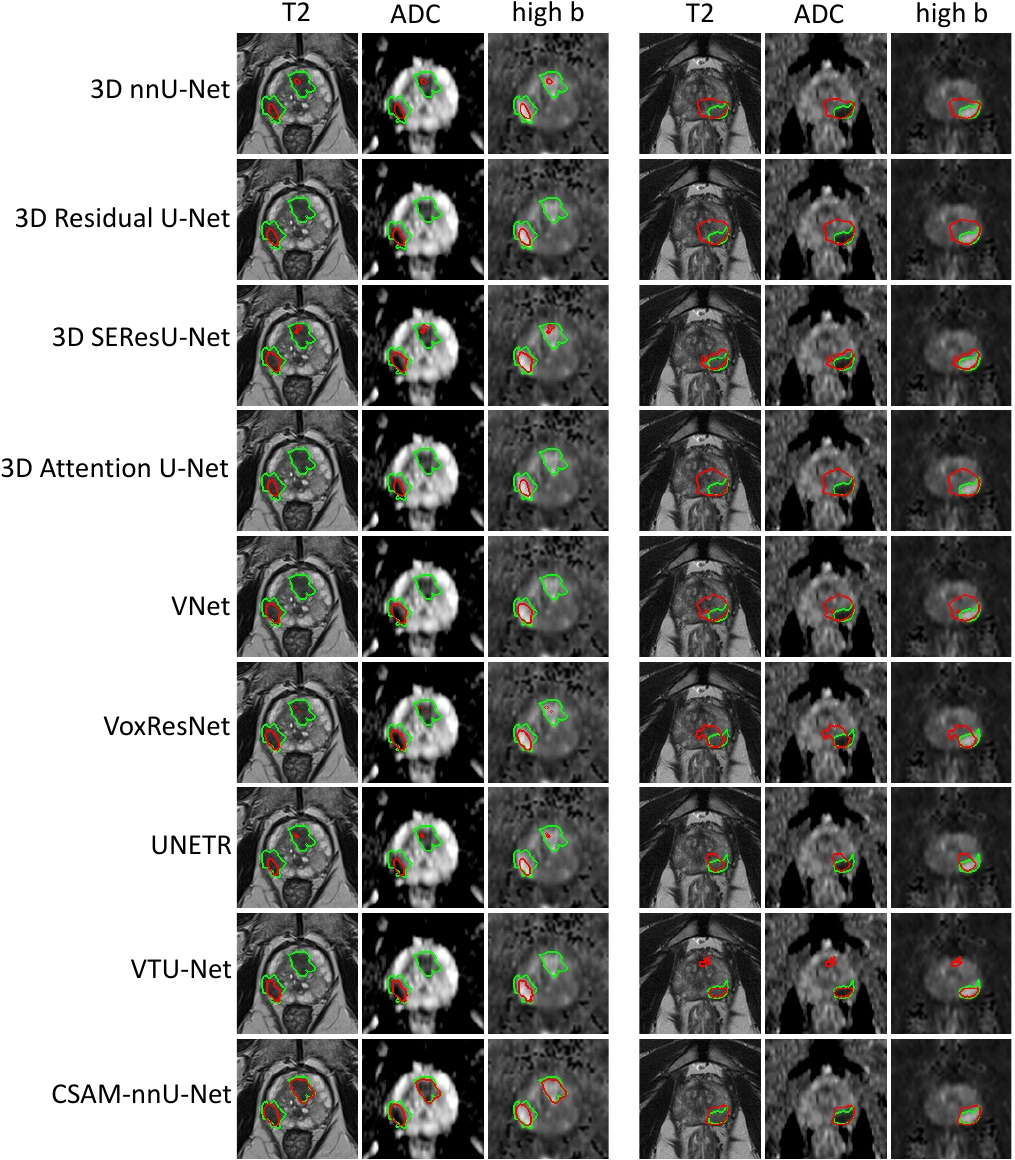}
     \caption{Two qualitative results of PCa segmentation. Green indicates the ground truth and red indicates the segmentation results by the corresponding models.}
     \label{fig:cancer_res}
 \end{figure}

We evaluated the performance of the different models by DSC and the classification AUC. The cancer lesion is small relative to the entire image and, therefore, a DSC value for a cancer lesion is expected to be lower than for zonal segmentation. Since 2/3 of the data are patients without cancer, we also measured the classification AUC. Regarding classification, we define that only segmentations with no segmented cancer are negative patients. We compared our CSAM-nnU-Net with other state-of-the-art 3D segmentation models, 3D nnU-Net~\cite{isensee2018nnu}, 3D Residual U-Net~\cite{Bhalerao2020ResUNet}, 3D SEResNet~\cite{hu2018squeeze}, 3D Attention U-Net~\cite{Oktay2018AttUNet}, VNet~\cite{milletari2016v}, VoxResNet~\cite{Chen2018VoxResNet}, UNETR~\cite{hatamizadeh2022unetr}, and VTU-Net~\cite{peiris2022robust}, and the results are reported in Table~\ref{tab:cancer_segmentation}. CSAM-nnU-Net performed the best in both segmentation and classification. The qualitative results are shown in Figure~\ref{fig:cancer_res}, which reveals that only CSAM-nnU-Net is able to segment both lesions in the left example, whereas other models all failed to segment the lesion on the right. Additionally, since the lesions appear to be slightly different among T2, ADC, and high-b images, the manual ground truth label fits the T2 image better while the segmentation by CSAM-nnU-Net fits the other images better. In the example on the right, other models either over-predicted the segmentation or generated segmentations with wrong lesion shapes, while the CSAM-nnU-Net result was closest to the ground truth.

\subsection{Placenta MRI Segmentation}

\begin{table}
    \centering
    \begin{tabular}{lll}
    \toprule
    &DSC (\%)$\uparrow$& RAVD (\%)$\downarrow$\\
    \midrule
         nnU-Net~\cite{isensee2018nnu}&81.2&36.5  \\
         3D nnU-Net~\cite{isensee2018nnu}&64.1&49.8 \\
         GT-nnU-Net~\cite{guo2021transformer}&82.0&34.8\\
         SAnnU-Net~\cite{zhang2020sau}&82.6&34.7\\
         CSAM-nnU-Net&\bfseries83.8&\bfseries32.5\\
    \midrule
         MSU-Net~\cite{su2021msu}&65.5&51.8  \\
         3D MSU-Net~\cite{su2021msu}&62.3&55.0 \\
         GT-MSU-Net~\cite{guo2021transformer}&63.6&54.9\\
         SAMSU-Net~\cite{zhang2020sau}&\bfseries69.4&53.7\\
         CSAM-MSU-Net&\bfseries69.4&\bfseries50.9\\
    \midrule
         nnU-Net++~\cite{zhou2018unet++}&65.6&53.7  \\
         3D nnU-Net++~\cite{zhou2018unet++}&70.7&46.4\\
         GT-nnU-Net++~\cite{guo2021transformer}&83.0&\bfseries34.2\\
         SAnnU-Net++~\cite{zhang2020sau}&62.8&52.9\\
         CSAM-nnU-Net++&\bfseries83.2&34.7\\
    \bottomrule
    \end{tabular}
    \caption{Placenta segmentation results (no statistical testing).}
    \label{tab:placenta_segmentation}
\end{table}

\begin{figure*}
    \centering
    \includegraphics[width=1\linewidth]{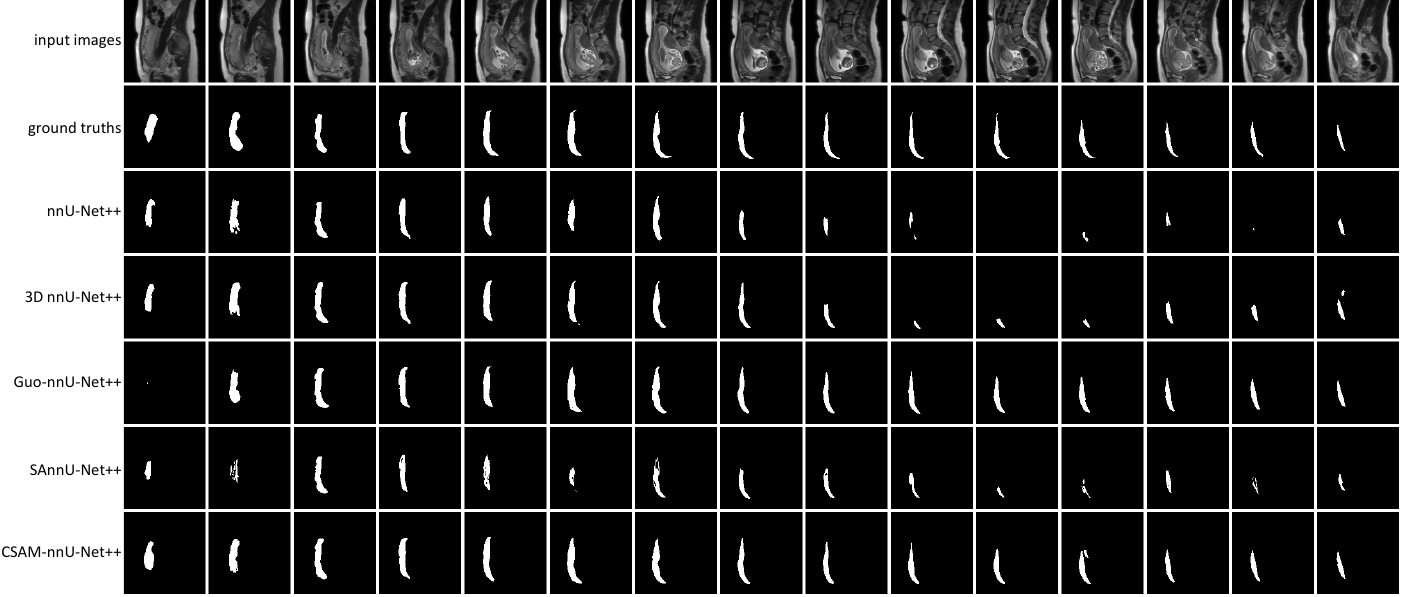}
  \caption{Results of placenta segmentation on consecutive slices. All the other nnU-Net++-based methods under-segment.
  }
  \label{fig:placenta_res}
  \end{figure*}


The data was acquired in three orthogonal planes (axial, sagittal, and coronal), including 14 to 18 weeks gestational age (GA), from eligible pregnant women~\cite{janzen2022description}. It includes 150 pregnancies, among which  105, 15, and 30 were used for training, validation, and testing, respectively. For training, we resized the volumes to $128\times128\times64$ for all three imaging views. 

We compared the CSAM-based models with the corresponding 2D and 3D models, as well as GT-based~\cite{guo2021transformer} and SA-based~\cite{zhang2020sau} 2.5D models on nnU-Net, MSU-Net, and nnU-Net++. As shown in Table~\ref{tab:placenta_segmentation}, our CSAM-based models are always the ones with the best performance, while other modules or methods are inconsistent on different backbones. For example, SAnnU-Net and SAMSU-Net are not too much worse than the corresponding CSAM-based models, but SAnnU-Net++ is the worst among all the competing nnU-Net++-based models. The GT method works fine on nnU-Net and nnU-Net++, but it is worse than the pure 2D model when it is applied on MSU-Net. An example of the segmentation by nnU-Net++-based models is shown in Figure~\ref{fig:placenta_res}, and we see that nnU-Net++, 3D nnU-Net++, and SAnnU-Net++ under-segment the placenta, while GT-nnU-Net++ fails to segment the placenta on the first slice. Our CSAM-based method yields the best result.

\subsection{Cardiac MRI Segmentation}


We utilized the ACDC public dataset~\cite{bernard2018deep}, which consists of 100 subjects (20 healthy subjects, 20 subjects with previous myocardial infarction, 20 subjects with dilated cardiomyopathy, 20 subjects with hypertrophic cardiomyopathy, and 20 subjects with abnormal right ventricle). The segmentation labels included right ventricular endocardium (RV), myocardium (MYO), and left ventricular endocardium (LV) for both the end diastole and end systole phases. We performed 5-fold cross-validation in this experiment.

The results are reported in Table~\ref{tab:ACDC_segmentation}.
Our CSAM showed a consistent improvement over the corresponding 2D and 2.5D models, and the CSAM-based models were always the best-performing methods. Even in some cases where the CSAM-based methods do not have the highest DSC, there is no statistical difference between the performance of the CSAM-based methods and the competing methods.

\begin{table}
\setlength{\tabcolsep}{2.5pt} 
    \centering
    \begin{tabular}{lllll}
    \toprule
    &RV&MYO&LV&Average\\
    \midrule
         nnU-Net~\cite{isensee2018nnu}&\bfseries86.8&\bfseries85.5&92.4&89.2\\
         GT-nnU-Net~\cite{guo2021transformer}&86.3&\bfseries85.5&\bfseries92.5&89.2\\
         SAnnU-Net~\cite{zhang2020sau}&86.2&85.1&92.2&88.9**\\
         CAT-nnU-Net~\cite{hung2022cat}&84.8*&84.0**&91.8&88.1** \\
         CSAM-nnU-Net&\bfseries86.8&\bfseries85.5&\bfseries92.5&\bfseries89.3\\
    \midrule
         MSU-Net~\cite{su2021msu}&82.9&83.6**&91.3&87.0*  \\
         GT-MSU-Net~\cite{guo2021transformer}&\bfseries83.8&83.6*&91.2&87.4\\
         SAMSU-Net~\cite{zhang2020sau}&82.4&82.4***&90.6**&86.3*\\
         CAT-MSU-Net~\cite{hung2022cat}&79.3**&81.3***&90.1**&85.3** \\
         CSAM-MSU-Net&83.3&\bfseries84.3&\bfseries91.8&\bfseries87.7\\
    \midrule
         nnU-Net++~\cite{zhou2018unet++}&83.0**&83.0***&90.3***&86.5***\\
         GT-nnU-Net++~\cite{guo2021transformer}&86.5&\bfseries85.7&92.5&\bfseries89.2\\
         SAnnU-Net++~\cite{zhang2020sau}&82.1**&81.6***&89.8***&85.6***\\
         CAT-nnU-Net++~\cite{hung2022cat}&86.2&84.7**&92.1*&88.9*\\
         CSAM-nnU-Net++&\bfseries86.6&85.4&\bfseries92.7&\bfseries89.2\\
    \bottomrule
    \end{tabular}
    \caption{Comparison on cardiac MRI segmentation (DSC in \%).}
    \label{tab:ACDC_segmentation}
\end{table}

\section{Conclusions}

For the segmentation for anisotropic volumetric MRI data, we have devised a 2.5D cross-slice attention module (CSAM) that introduces minimal additional parameters to train, which can be incorporated into a variety of segmentation network backbones. Our extensive experiments showed that, despite the relatively small number of additional parameters, our CSAM-based models outperform the associated baseline 2D, 3D, and 2.5D models under simple training and data augmentation settings consistently across different segmentation tasks. Our study confirms that the CSAM's ability to learn and leverage cross-slice information within 3D image volumes improves volumetric segmentation performance over 2D and 3D methods and that our models generalize better than the state-of-the-art 2.5D modules.

Our research opens up promising avenues for future work. In particular, since pooling is performed globally in CSAM, it may be possible to improve performance and generalizability by incorporating into the cross-slice attention strategy information that is more local. Additionally, a further study of uncertainty measurement is needed, as this is critical to understanding the limitations of deep learning-based medical image segmentation models, which would help clinicians make better informed decisions.

{\small
\bibliographystyle{IEEEtranS}
\bibliography{egbib}
}

\end{document}